\documentclass[aps,prb,
twocolumn,floatfix]{revtex4}

\usepackage{epsfig}
\usepackage{graphicx}
\usepackage{amsfonts}

\begin{document}

\title{ Formation of a Quasi-Periodic Copper Thin Film}

\author{J. Ledieu, J.-T. Hoeft, D.E. Reid and J. Smerdon}
\affiliation{Surface Science Research Centre,
	The University of Liverpool, Liverpool L69 3BX, UK}
\author{R.D. Diehl}

\affiliation{Department of Physics,
	Pennsylvania State University,
	University Park, PA 16802, USA}
\author{ T.A. Lograsso and A.R. Ross}
\affiliation{Ames Laboratory, Iowa State University,
	Ames, IA 50011, USA}
\author{R. McGrath}
\affiliation{Surface Science Research Centre and
Department of Physics, The University of Liverpool,
Liverpool L69 3BX, UK}

\begin{abstract}
We have synthesised a thin film of copper with a quasi-periodic 
structure by the adsorption of copper atoms on the five-fold surface 
of the icosahedral quasicrystal Al-Pd-Mn at room temperature.  The 
quasi-periodicity of the thin film is manifested in low energy 
electronic diffraction (LEED) measurements and in the existence of 
Fibonacci relationships between rows of copper atoms imaged using 
scanning tunneling microscopy (STM).  These findings demonstrate the 
feasibility of single-element quasi-periodic thin film formation using 
quasicrystals as templates.
\end{abstract}

\pacs{61.44.Br, 68.35.Bs, 68.37.Ef, 61.14.Hg}

\maketitle

Copper is one of the oldest elements known to mankind and it 
crystallizes in a face-centred cubic structure.  Quasicrystals are a 
relatively new form of matter, first reported in 1984 \cite{Shechtman84}, and are bi- 
and tri-metallic alloys with long-range order but no translational 
symmetry.  In the course of investigations aimed at understanding the 
interactions between quasicrystal surfaces and adsorbed atomic and 
molecular species \cite{McGrath02}, we have discovered that the adsorption of Cu on 
the five-fold surface of the icosahedral quasicrystal Al-Pd-Mn at 300 
K leads to the formation of a quasi-periodic Cu thin film.  The 
quasi-periodic structure of the thin film is manifested in low energy 
electronic diffraction (LEED) measurements and in the existence of 
Fibonacci relationships between rows of Cu atoms on the surface imaged 
using scanning tunneling microscopy (STM).  The ability to synthesise 
such single-element quasi-periodic thin films will facilitate the 
study of the relationship between quasi-periodicity and physical 
properties and enable the probing of the transition from 
two-dimensional (2D) to three-dimensional (3D) electronic properties 
in a quasi-periodic material.

The starting point in these investigations is the preparation of high 
quality clean surfaces of Al-Pd-Mn with large flat terraces (microns 
in size) and low surface corrugation in an ultra-high-vacuum (UHV) 
environment.  The methodology for the preparation of these surfaces 
has been previously described in detail \cite{Ledieu01,Papadopolos02} and consists of ex-situ 
polishing followed by several cycles of ion sputtering (~1 hour) and 
annealing to 940 K ($\approx$4 hours) with a cumulative anneal time of ~20 
hours.  Surface preparation is facilitated using LEED to establish the 
degree of surface ordering and Auger electron spectroscopy (AES) to 
check for contamination (and later to monitor Cu deposition).  
Surfaces prepared in this manner have been shown to be essentially 
ideal terminations of the bulk quasicrystal structure 
\cite{Ledieu01,Barbier02}, and the 
2D structure of these surfaces has been described as a ÒFibonacci 
pentagridÓ \cite{Schaub94a}, with the five principal symmetry axes evident in the 
arrangement of the dominant structural motifs.  This is illustrated in 
Fig.  1(a) which shows a high resolution STM image of a small area (10 
nm x 10 nm) of such a surface.  Several groupings of atoms with 
five-fold and ten-fold symmetry are evident.  The dark five-fold 
ÒstarsÓ are Bergman clusters (a basic structural entity of this type 
of quasicrystal) which have been truncated in the surface formation 
process \cite{Papadopolos02}.  White lines drawn on the figure join identical positions 
on the dark five-fold stars; the separations between the lines are of 
two lengths $L = 0.74 \pm 0.02$ nm and $S = 0.46 \pm 0.02$ nm, whose ratio is 
within experimental error equal to the golden mean $\tau$; [$\tau$ is the 
irrational number expected for the ratio of the two basic units of the 
Fibonacci sequence [$\tau=(\sqrt{5}+1)/2=1.618...$].  Furthermore the arrangement 
of these line separations (reading from bottom to top) $LSLLSL$ forms a 
segment of the Fibonacci sequence.

\begin{figure}[bh]
\begin{center} 
\epsfxsize=75mm
\epsffile{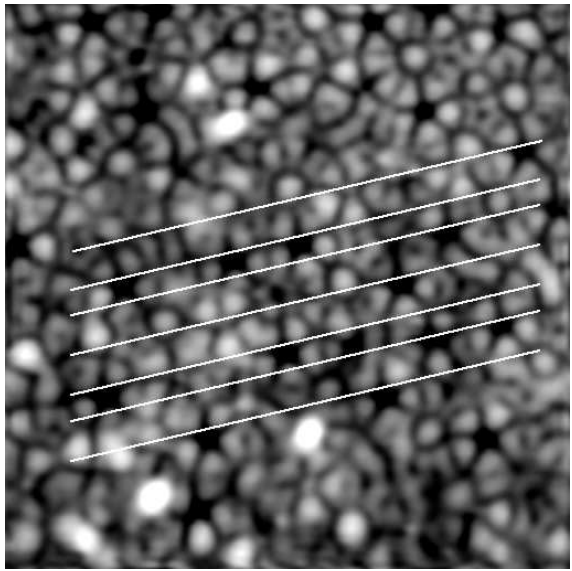}
\caption{
10 nm x 10 nm STM image of the five-fold surface of Al-Pd-Mn.  The 
lines join identical positions on the dark five-fold stars; the 
separations between the lines are of two lengths $L = 0.74 \pm 0.02$ nm 
and $S = 0.46 \pm 0.02$ nm, whose ratio is within experimental error equal 
to the golden mean $\tau$; furthermore the arrangement of these line 
separations (from bottom to top) $LSLLSL$ forms a segment of the 
Fibonacci sequence.
}
\label{clean}
\end{center}
\end{figure}

Deposition of Cu atoms on the surface was achieved using a simple 
evaporation source consisting of a tantalum filament which was wrapped 
around a sample of pure Cu.  The filament was thoroughly degassed 
before evaporation and the evaporation flux was found to be very 
consistent under constant conditions of the control parameters.  The 
experiments were conducted using a Cu flux of $4.5 \pm 0.2 x 10^{-2}$ 
monolayers s$^{-1}$ as determined by measuring the fractional area of the 
surface covered with successive Cu depositions.  The sample was at 
room temperature during deposition and measurement.

Figure 2 gives an overview of the growth of Cu on the surface at room 
temperature for increasing Cu coverage.  Under the above deposition 
conditions the growth of Cu on the Al-Pd-Mn surface is observed to 
proceed in a layer-by-layer manner.  The step height from layer to 
layer is $0.19 \pm 0.01$ nm.  For sub-monolayer coverages (Fig.  2(a)) Cu 
islands form on the surface and as adsorption continues the second 
layer is observed to begin growing when the first layer is $90 \pm 5$ \% 
complete.  For layers 1 to 4 the Cu atoms within the layers are 
observed to self-organise into small domains within which there is a 
developing pattern of one-dimensional rows (Fig 2(b)).  From layers 5 
to 8 a very well-developed row structure is observed (Fig.  2(c)).  
From layer 9 the effect of layer growth before previous layer 
completion becomes more marked as shown in Fig.  2(d); this results in 
a diminishing domain size and in the simultaneous existence of many 
incomplete layers at the surface.  The row structure persists in these 
layers.  The results were reproduced several times and the growth was 
monitored up to the 25th.  layer.

\begin{figure}[bh]
\begin{center}
\epsfxsize=75mm
\epsffile{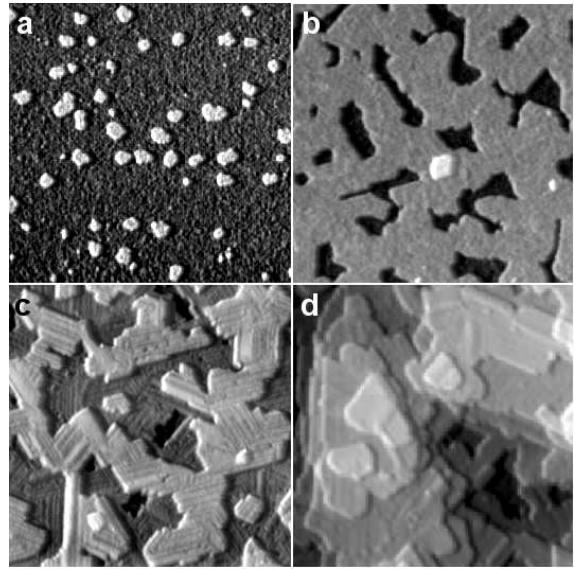}
\caption{(a) 50 nm x 50 nm STM snapshots of the five-fold surface of 
Al-Pd-Mn during the growth of the Cu thin film.  (a) 0.09 layers of 
Cu, (b) 3.8 layers, (c) 5.5 layers, (d) 11.7 layers.
}
\label{figure2}
\end{center}
\end{figure}

Fig.  3(a) is a 40nm x 40 nm image taken during the growth of the 6th 
layer.  This image reveals several intriguing details.  The islands 
have sharp edges where the border is a single row.  A plausible 
explanation for the growth mechanism is the initial adsorption of Cu 
on top of a growing layer, followed by diffusion of the adsorbed atoms 
along the Cu rows and subsequent attachment at the end of each row.  
The rows themselves appear in five orientations, with the angles made 
by intersections of rows in the same layer corresponding to the 
internal angles of a pentagon.  The intersections between rows are of 
two types.  Two sets of rows can meet at vertex points, where the end 
of each row in one direction meets the end of another at an angle of 
108$^{o}$ (the internal angle of a pentagon).  Alternatively the rows in 
one direction intersect a single row at an angle of 72$^{o}$.  The rows do 
not form a periodic structure.

\begin{figure}[bh]
\begin{center}
\epsfxsize=75mm
\epsffile{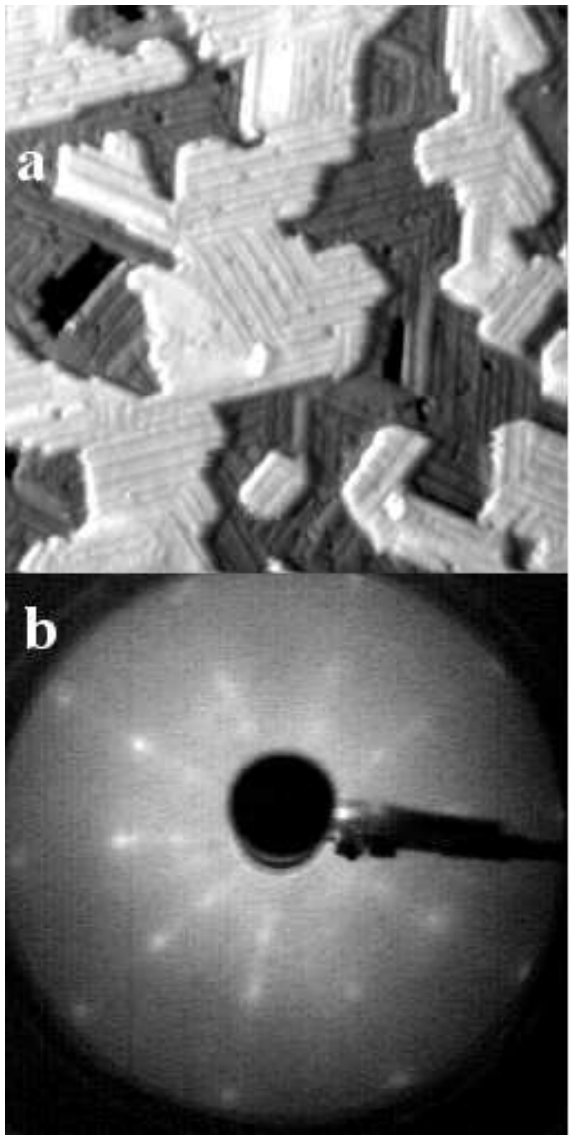}
\caption{(a) 40 nm x 40 nm STM image of the five-fold surface of 
Al-Pd-Mn after deposition of 5.5 ML of Cu.  (b) LEED pattern (beam 
energy 50 eV) corresponding to this phase.  The relationships between 
spot positions are indicative of $\tau$-scaling within experimental error.}
\label{fig3}
\end{center}
\end{figure}

The corrugation across the rows is measured at $0.025 \pm 0.005$ nm using 
STM.  The atomic structure within the rows is not resolved in these 
experiments; this could be an indication of a large vibrational 
amplitude along the rows; alternatively it could reflect a low 
corrugation in electronic density along the rows.  The fact that the 
LEED pattern shows discrete spots is indicative of long-range order 
both within and across domains and along the rows themselves.  The row 
directions are also observed to be correlated from layer to layer, 
indicating that each layer acts as a template for the subsequent one 
in the growth process.

The inter-row distances form aperiodic sequences having long and short 
separations.  This is illustrated in Figure 4 which shows a 10 nm x 10 
nm image of the surface during formation of the 6th layer.  There are 
some vacancy defects in the rows themselves.  Nevertheless sequences 
of rows having S and L separations are visible.  The average spacings 
are measured at $0.45 \pm 0.02$ nm and $L = 0.73 \pm 0.03$ nm.  The ratio of 
these numbers equals the golden mean $\tau$ within experimental error.  Two 
such sequences ($SLLSLSL$, reading from bottom to top) are indicated on 
this figure, on rows which meet at vertex points.  This indicates that 
the same sequencing information is common to both sets of rows.  The 
coherence length of the row structures appears to be determined by the 
domain walls which arise during the growth process.  The close match 
of these row spacings in the Cu structure with those illustrated for 
the clean surfaces (Figure 1) indicate that the clean surfaces acts as 
a template for the ordered growth of the Cu atoms.

\begin{figure}[bh]
\begin{center}
\epsfxsize=75mm
\epsffile{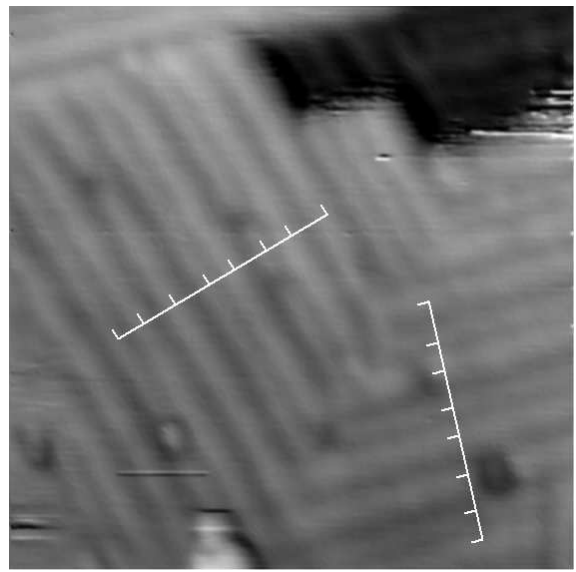}
\caption{(a) 10 nm x 10 nm STM image of the five-fold surface of 
Al-Pd-Mn.  The lines marks sequences of Cu rows with spacings given by 
$SLLSLSL$ (from bottom to top).  The spacings are $S= 0.45 \pm 0.02$ nm and 
$L = 0.73 \pm 0.03$ nm.  The ratio of these numbers equals the golden mean 
$\tau$ within experimental error.}
\label{fig4}
\end{center}
\end{figure}

The LEED pattern characteristic of this layered row structure is shown 
in Fig.  3(b).  The pattern has ten-fold symmetric rings of spots 
whose distances from the central spot (hidden by the electron gun in 
the Figure) exhibit a $\tau$-scaling relationship.  From layers 9 to 20, 
with the simultaneous growth of multiple layers, the LEED pattern 
becomes streaky and diffuse and eventually degrades and disappears by 
the 25th layer.  This is consistent with the diminishing domain size.

Upon annealing to 570 K, STM images (not shown) reveal that the Cu 
thin film undergoes an irreversible transformation to a cubic 
structure Cu with five domains rotated from each other by 72$^{o}$.  
The LEED pattern associated with this phase, although 10-fold 
symmetric, no longer has the $\tau$-scaling relationships in the distances 
of the diffraction spots.  The domain boundaries of this phase are 
decorated with excess Cu atoms.  Flashing the sample to 660 K results 
in thermal desorption of the Cu film and the clean surface LEED 
pattern is restored.

There has been a report of the use of these surfaces for the growth of 
five-fold symmetric nanoclusters \cite{Cai03} by the adsorption of 
sub-monolayer amounts of aluminum on Al-Cu-Fe.  There has been a 
report of ordered atomic Si and Bi monolayers on quasicrystal surfaces 
\cite{Franke02}, and another report of the formation of a quasicrystalline 
bi-metallic Au-Al alloy by deposition and subsequent annealing of a Au 
film on an Al-Pd-Mn substrate \cite{Shimoda01}.  In other work, the deposition of 
Ag at low temperature onto to a GaAs(110) substrate followed by 
annealing to room temperature was found to form a Ag (111) film with a 
surface quasiperiodic modulation \cite{Smith96, Ebert99}.  However the system 
described in this report is unique in that it constitutes a single 
element quasi-periodic thin film grown using a quasicrystal surface as 
a template.

 The quasi-periodicity of the film, together with the 
 one-dimensionality of the row structure, suggest the possibility of 
 unusual vibrational and electronic properties in this system.  The 
 well-documented low density of electronic states at the Fermi level in 
 the bulk quasicrystal (the so-called ÒpseudogapÓ) means features in 
 the valence electronic structure of the Cu film should be readily 
 distinguishable from those of the substrate.  This system also offers 
 interesting possibilities for monitoring the transition from 2D to 3D 
 electronic properties in a quasi-periodic material as the thickness of 
 the film is increased, and for probing the relationship between 
 quasi-periodicity and physical properties.  The surface of the film 
 itself has the potential to act as an adsorption template for the 
 formation of further quasi-periodic systems.  Furthermore we have no 
 reason to believe that the Cu/Al-Pd-Mn system is unique; we anticipate 
 the discovery of further aperiodic thin film systems, with the promise 
 of new electronic and/or magnetic properties.

\textbf{Acknowledgements} 
The UK Engineering and Physical Sciences Research Council, the US 
National Science Foundation, the U.S.  Department of Energy, Basic 
Energy Sciences, and the EU Marie Curie Host Development Fund are 
acknowledged for funding.  V.  Humblot is thanked for experimental 
assistance and N.  Brabner for the construction of the Cu evaporator.
\bibliography{copper}
\bibliographystyle{prsty}

\end{document}